\documentclass[final,5p,times,twocolumn]{elsarticle}

\usepackage{amssymb}

\journal{Physica C}

\begin{document}

\begin{frontmatter}



\title{Negative nonlocal and local voltages (resistances) in a quasi-one-dimensional superconducting aluminum structure}


\author{V.~I.~Kuznetsov \corref{cor1}}
\ead{kuznetcvova@mail.ru}
\cortext[cor1]{corresponding author}
\author{O.~V.~Trofimov}
\address{Institute of Microelectronics Technology and High
Purity Materials, Russian Academy of Sciences, Chernogolovka,
Moscow Region 142432, Russia}

\begin{abstract}
To study a nonlocal electron transport in an aluminum
superconducting quasi-one-dimensional structure, we measured
negative nonlocal (local) direct current voltages in the structure
in a magnetic field near the critical temperature. The structure
is a normal-superconducting at $T_{cn}<T<T_{cw}$ ($T_{cn}$ and
$T_{cw}$ are the critical temperatures for narrow and wide wires,
respectively, making up this structure). Negative voltage arises
due to a quasiparticle current flowing through the N-S interface.
We plotted the experimental and theoretical temperature and
magnetic-field dependences of current, resistance and voltage
corresponding to the peak of negative voltage, taking into account
either equilibrium or nonequilibrium superconducting fluctuations.
\end{abstract}




\end{frontmatter}


\section{INTRODUCTION}
Nonlocal phenomena in superconducting and hybrid nanostructures
consisting of a normal (N) metal (magnetic (F) material, a
semiconductor wire) and a superconductor (S) are not studied well.
We understand the term "nonlocality" as the effect of electron
transport in one part of the structure on electron transport in
other parts of the structure. The study of nonocal phenomena is of
great interest due to the miniaturization of microdevices and the
development of devices using nonlocal effects.

Nonlocal voltage (resistance) appears in those parts of the hybrid
structures through which the applied current does not flow, due to
crossed Andreev reflection (CAR), the elastic co-tunneling  (EC)
and quasiparticle charge imbalance (QCI) \cite{caddenjiang}.

Nonlocal processes have different spatial scales. The EC, CAR
processes operate on a shorter scale, close to the
temperature-dependent superconducting coherence length $\xi(T)$.
Whereas, the nonequilibrium process QCI appears on a sufficiently
long scale close to the length of the quasiparticle diffusion
$\lambda_{Q}(T,B)$ dependent on a temperature and a magnetic field
(relaxation length of the quasiparticle charge imbalance)
\cite{schmidt}.

An interesting nonlocal non-equilibrium effect was found in a
mesoscopic SNS junction with an additional N probe
\cite{golikova}. A quasiparticle current was injected into one of
the S banks of the SNS junction through the N probe, located
outside the central part of the SNS junction. In this case, the
nonlocal voltage between the S probes was measured in the central
part of the SNS junction, through which the applied current did
not flow. The nonlocal voltage was zero until the injected
quasiparticle current reached a certain critical value. This
effect is due to the appearance of nonlocal quasiparticle and
superconducting currents flowing in the central part of the SNS
junction. This nonlocal superconducting current arose as a
countercurrent, compensating for the nonlocal quasiparticle
current. When the nonlocal superconducting current exceeded its
critical value, a non-zero nonlocal voltage appeared.

Earlier, new nonlocal nonequilibrium effects related to the
quasiparticle charge imbalance were found in the electron
transport of both superconducting quasi-one-dimensional aluminum
wires with single and variable widths \cite{kuznjetplet16}, as
well as single \cite{dubjetplet03, kuznjetplet19} and double
\cite{kuznprb08, kuznphysica13} thin-film loops, composed of such
wires, biased with a direct current or an alternating current
(without a direct current component) and penetrated by a magnetic
flux $\Phi$ at temperatures $T$ slightly below the critical
temperature $T_{c}$.

Circular-asymmetric aluminum single loops and a series of such
identical single loops \cite{dubjetplet03} are a highly effective
alternating voltage rectifier. In this case, the rectified voltage
oscillates as a function of a field $B$ with a period of
$dB=\Phi_{0}/S_{L}$ (where $\Phi_{0}=hc/2e$ is the superconducting
magnetic flux quantum, $S_{L}$ is an effective loop area). The
total rectified voltage on a series of $n$ loops is $n$ times
greater than this voltage on one loop. Two circular-asymmetric
loops of different areas directly connected \cite{kuznprb08} or
making up a series of two loops \cite{kuznphysica13} show nonlocal
long-range coupling between the quantum electron transport of one
loop and the transport of the other loop.

Plateaus of an almost constant dc voltage were measured at
voltages $V_{pl}=2\Delta(T,B)/ne$ (where $\Delta(T,B)$ is a
temperature and field-dependent superconducting gap taken out of
the center of the nonequilibrium region, $n$ is an integer, $e$ is
the electron charge) on the $V(I)$ curves of quasi-one-dimensional
superconducting aluminum wires \cite{kuznjetplet16}. Plateaus
arise due to multiple Andreev reflection with a large number of
reflections (up to $n=32$) and strong quasiparticle overheating
that occur in a self-formed short SNS junction or a phase-slip
center (PSC).

Unexpected quantum fractional $\Phi_{0}/m=(hc/2e)/m$ periodic
($m=2-20$) oscillations of voltage $V(B)$ as a function of $B$
have been experimentally studied in an aluminum superconducting
loop penetrated by a magnetic flux under strongly nonequilibrium
conditions (applied direct current much higher than the critical
current and $T<T_{c}$) \cite{kuznjetplet19}. Decreasing in the
oscillation period by a factor of $m$ relative to the value of
$\Phi_{0}$ is interpreted as increasing in the effective charge of
Cooper pairs by a factor of $m$ due to multiple Andreev reflection
in the SNS junction or PSC formed in the loop.

In order to clarify the features and mechanisms of nonequilibrium
nonlocal and local effects of \cite{kuznjetplet16, dubjetplet03,
kuznjetplet19, kuznprb08, kuznphysica13}, we have experimentally
investigated nonlocal and local nonequilibrium electron transport
in the aluminum superconducting quasi-one-dimensional structure
(inset of  Fig.  \ref{f1}), placed in the field $B$, perpendicular
to the substrate surface, at $T$ very close to $T_{c}$. We have
recorded the nonlocal $V_{NL}(I)$ and local $V_{L}(I)$ voltages on
the structure biased with a direct current $I$ using different
measurement circuits.

\section{EXPERIMENTAL PROCEDURE AND STRUCTURE}
For measuring low voltage (less than one $\mu$V) as a function of
dc current applied to a superconducting microstructure, protection
is necessary from low and high-frequency electromagnetic
interference and noise. In this work, electrical measurements were
carried out in a room shielded from high-frequency interference,
at that time of day when other electrical installations and
devices located in other rooms were not working. The absence of
other electrical installations made it possible to minimize the
influence of power-supply and low-frequency interferences. In
addition, almost all measurements were carried out using analog
devices, since the use of digital devices could lead to distortion
and smearing of weak signals. Most of the devices used were
designed in our laboratory and powered by galvanic batteries. The
peak-to-peak output voltage divided by the voltage gain of 1000 on
our low-frequency dc amplifier, with the amplifier input shorted,
was 0.03 $\mu$V. The voltage from the structure through this
amplifier was applied to the Y coordinate block of the plotter. A
home-made low-frequency generator was also taken for the dc sweep.
The voltage proportional to the external current was applied to
the X coordinate block of the plotter. Most of $V(I)$ curves were
recorded at a current sweep rate of 0.045 $\mu$A/s. Decreasing
(increasing) the sweep rate by several times did not significantly
affect the measurement result. The wiring inside the cryostat was
made using twisted pairs of copper wires 0.15 mm in diameter. The
low-temperature end of each copper wire was soldered in series
with a single 1 k$\Omega$ resistance located on a cold chip holder
and then connected in series with the electrical wiring on the
chip with the structure. These resistances serve to reduce the
impact of noise and to protect the structure under study from
electrical breakdown. The electrical measurements presented in
this work were carried out, for the most part, without using other
filters, since great measures were taken to minimize the impact of
electromagnetic interference on the structure.

\begin{figure}
\begin{center}
\includegraphics[width=1\linewidth]{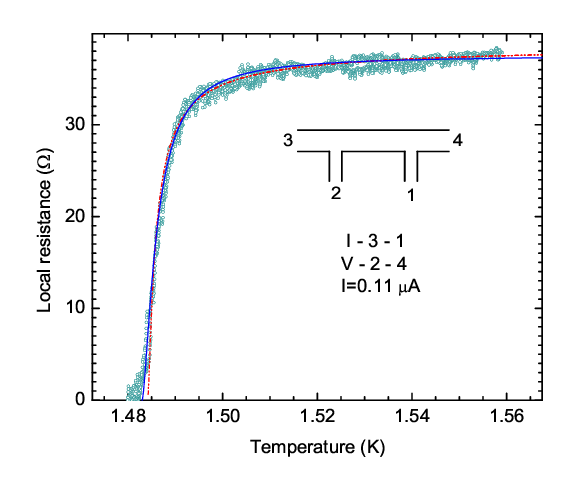}
\caption{\label{f1} (Color online) Resistive N-S transition
$R_{0}(T)$ as a function of $T$ at $I=0.11$ $\mu$A. Circles are
experimental data. The solid and dash-dotted lines are fitting
functions $R_{1}(T)$ and $R_{2}(T)$, taking into account one
$\sigma^{1AL}$ correction to the conductivity  and two
$\sigma^{1AL}$ and $\sigma^{1AMT}$ corrections, respectively.
Inset: a sketch of the structure.}
\end{center}
\end{figure}

A sketch of the central part of the studied structure is shown in
the inset of Fig. \ref{f1}. The structure was obtained by the
thermal deposition of aluminum with the film thickness $d=19$ nm
on a silicon substrate using the lift-off process of electron beam
lithography. The structure consists of the wide wire (3-4) with
the width $w_{w}=0.50$ $\mu$m and adjacent narrow wires (2 and 1)
with the width $w_{n}=0.27$ $\mu$m. The structure is
quasi-one-dimensional, since $d<<w_{n}<w_{w}<2\xi(T)$ in the
studied temperature range. The distance between wires 2 and 1,
equal to $L_{21}=6.69$ $\mu$m, satisfies the requirement
$\xi(T)<<L_{21}\approx \lambda_{Q}(T,B)$.

Together with the structure under study, a reference narrow wire
with a width of $w_{n}=0.27$ $\mu$m was deposited in one cycle on
one chip. It should be expected that the corresponding parameters
of the reference narrow wire and the narrow wire of the structure
under study are similar. In addition, two other aluminum
homogeneous narrow and wide wires with the thickness of $d
\backsimeq 19$ nm and widths of $w_{n} \backsimeq 0.27$ and $w_{w}
\backsimeq 0.5$ $\mu$m were fabricated on one chip using the same
technology.

The structure has the following parameters: the resistance of the
wire with the length of $L_{21}$ in the normal state is
$R_{0}=37.85$\,$\Omega$, the ratio of resistances at temperatures
of 300 and 4.2 K $R_{300}/R_{4.2}=2$, the resistance per square in
the normal state $R_{sq}=2.83$\,$\Omega$, the resistivity of the
wire $\rho_{n}=5.38\times 10^{-8}$ $\Omega$\,m. The critical
temperature $T_{c}=1.486$\,K is determined at the middle of the
resistive N-S transition $R_{0}(T)$ at the direct current of 0.11
$\mu$A. The mean free path of the electron $l=9.5$ nm was obtained
from the refined theoretical expression for aluminum
\cite{gershenson} $\rho_{n}l=5.1\times 10^{-16}$
$\Omega$\,m$^{2}$. The electron diffusion coefficient
$D=v_{F}l/3=4.1\times 10^{-3}$ m$^{2}$/s, where $v_{F}
=1.3\times10^{6}$ m/s is the Fermi velocity for aluminum. The
inelastic electron - phonon scattering time for aluminum
\cite{stuivinga} is $\tau_{e-p}=1.3\times10^{-8}$\,s at
$T=1.2$\,K. Assuming the temperature dependence of the
electron-phonon scattering time $\tau_{e-p}(T) \propto T^{-3}$
near $T_{c}$ \cite{kaplan}, we have recalculated this time for
temperatures $T_{c}=1.486$ and $T=1.457$\,K and have obtained
$\tau_{e-p}(1.486)=6.8\times10^{-9}$ and
$\tau_{e-p}(1.457)=7.3\times10^{-9}$\,s, respectively.

The structure is a dirty superconductor, since $l<<\xi_{0}$, where
$\xi_{0}=1.6$ $\mu$m is the superconducting coherence length of
pure aluminum at $T=0$\,K. Then, near $T_{c}$, the Ginzburg-Landau
(GL) coherence length \cite{schmidt}
$\xi(T)=\xi(0)(1-T/T_{c})^{-1/2}$, where
$\xi(0)=0.85(l\xi_{0})^{1/2}$ is the coherence length at $T=0$\,K.
The temperature-dependent penetration depth of the magnetic field
\cite{schmidt} is $\lambda_{GL}(T)=\lambda(0)(1-T/T_{c})^{-1/2}$,
where $\lambda(0)=0.615\lambda_{L}(\xi_{0}/l)^{1/2}$ is the field
penetration depth at $T=0$\,K, $\lambda_{L}=16$ nm is the London
penetration depth for aluminum. For our structure, $\xi(0)=0.105$
$\mu$m, $\lambda(0)=0.128$ $\mu$m. The density of the
Ginzburg-Landau depairing critical current \cite{tinkham}
$j_{GL}(T)=j_{GL}(0)(1-T/T_{c})^{3/2}$, where
$j_{GL}(0)=c\Phi_{0}/12\sqrt{3}\pi^{2}\lambda(0)^{2}\xi(0)=1.01\times10^{-10}/\lambda(0)^{2}\xi(0)=5.9\times10^{10}$
A/m\,$^{2}$. Then, it should be expected that the GL depairing
critical currents at $T=0$ of narrow and wide wires (making up the
structure) with widths $w_{n}=0.27$ and $w_{w}=0.5$ $\mu$m will be
equal to $I_{GL}(0, w_{n})=300$ $\mu$A and $I_{GL}(0, w_{w})=560$
$\mu$A, respectively.

\section{RESULTS AND DISCUSSION}
\subsection{Resistive N-S transition $R_{0}(T)$}

The resistive N-S transition $R_{0}(T)$ of the central part of the
structure was recorded when an applied direct current $I=0.11$
$\mu$A was passed through probes 3 and 1, and the voltage $V$ was
taken using probes 2 and 4 (Fig. \ref{f1}). The inset of Fig.
\ref{f1} presents a sketch of the structure. Circles are data
measured in two cycles. The critical temperature $T_{c} =1.486$ K
is found at the middle of the N-S transition. In addition, we
found that the critical temperatures of the structure are equal to
1.486 and 1.484 K, when the $ R_{0}(T)$ transition is recorded at
a direct current $I = 0.11$ $\mu$A in accordance with the
measurement circuits ($I$ - 2-1, $V$ - 3-4) and ($I$ - 3-4, $V$ -
2-1), respectively. The solid and dash-dotted lines are the
fitting functions $R_{1}(T)$ and $R_{2}(T)$, respectively, taking
into account the additional conductivity within the framework of
the theory of superconducting fluctuations above $T_{c}$
\cite{larkin}.

The function $R_{1}(T)$ takes into account only the
Aslamazov-Larkin correction $\sigma^{1AL}$ \cite{larkin} to the
conductivity of a superconducting quasi-one-dimensional wire with
a cross section $A=w_{w}d$. The $R_{2}(T)$ function takes into
account two corrections to the conductivity: the Aslamazov-Larkin
(AL) correction $\sigma^{1AL}$ and the anomalous Maki-Thompson
(MT) correction $\sigma^{1AMT}$ \cite{larkin}.

Note that to plot $R_{2}(T)$ function, we used the expression for
$\sigma^{1AMT}$ given in \cite{larkin}. In addition, to plot some
other fitting function, another original expression for
$\sigma^{1AMT}$ was taken, obtained in \cite{thompson}. Fit the
experimental data with a function that takes into account the
$\sigma^{1AL}$ correction and the original expression for the
Maki-Thompson correction \cite{thompson}
$\sigma^{1AMT}=\sigma^{1AL}(-4+(8/\pi)/(\sqrt{\gamma}\tau))$ gives
the poor approximation, therefore, it is not presented here.

To calculate AL and MT corrections, we used the expression
\cite{larkin}  $\sigma^{1AL}=\sigma^{1AL}(0)\tau^{-3/2}$ (where
$\sigma^{1AL}(0)=e^{2}\pi\xi(0)/16\hbar A$, $\tau=T/T_{cf}-1$ -
normalized temperature difference, $T_{cf}$ is the fitting
critical temperature) and expression \cite{larkin}
$\sigma^{1AMT}=8\sigma^{1AL}(\tau/\gamma)/(1+\sqrt{\tau/\gamma})$
(where $\gamma=\pi\hbar/8kT\tau_{\varphi}$ is a dimensionless
depairing factor, $\tau_{\varphi}$ is a effective phase - breaking
time), respectively.

We plotted the fitting function
$R_{1}(T)=R_{0f1}/(1+C_{1}\rho_{n}\sigma^{1AL})=R_{0f1}/(1+C_{1}\rho_{n}\sigma^{1AL}(0)(T/T_{cf1}-1
)^{-3/2})$, where $R_{0f1}=37.55$ $\Omega$ and $T_{cf1}=1.483$ K
are the fitting resistance and critical temperature close to the
experimental values, $C_{1}=3.5$ is the factor characterizing the
deviation of the experimental value from the theoretical value of
the correction $\sigma^{1AL}(0)$.

Here, another fitting function is given
$R_{2}(T)=R_{0f2}/(1+\rho_{n}(\sigma^{1AL}+\sigma^{1AMT}))=R_{0f2}/(1+\rho_{n}\sigma^{1AL}(0)\tau^{-3/2}(1+8(\tau/\gamma_{f2})/(1+\sqrt{\tau/\gamma_{f2}})))$,
where $\tau=T/T_{cf2} -1$, $R_{0f2}=38.7$ $\Omega$,
$T_{cf2}=1.484$ K, and $\gamma_{f2}=0.011$ are the fitting
resistance, critical temperature and depairing factor,
respectively. Note that the fitting $T_{cf1}$ and $T_{cf2}$
correspond to the bottom of the resistive transition. The fitting
depairing factor $\gamma_{f2}$ is 6.9 times greater than the
calculated depairing factor $\gamma=0.0016$.

Depairing factor $\gamma=0.0016$ was calculated for a wire
$w_{w}=0.5$ $\mu$m at $T=1.486$ K using the total phase-breaking
frequency of the electron wave function
$\tau_{\varphi}^{-1}=\tau_{N}^{-1}+
\tau_{e-p}^{-1}=8.1\times10^{8}$ s$^{-1}$ (where
$\tau_{N}^{-1}=(\sqrt{D}e^{2}R_{sq}kT/\sqrt{2}w\hbar ^{2})^{2/3}$
\cite{altshuler} and $\tau_{e-p}^{-1}=1.5\times10^{8}$ s$^{-1}$ -
the phase-breaking frequencies due to an electron-electronic
interaction and due to inelastic electron-phonon scattering in a
quasi-one-dimensional wire, respectively). The large value of
$\gamma_{f2}$ may be due to additional unaccounted mechanisms for
phase coherence breaking, for example, external electromagnetic
noise.

\subsection{Nonlocal $V_{NL}(I)$ and local $V_{L}(I)$ curves at different measurement circuits at $T=1.462$\,K}

\begin{figure}
\begin{center}
\includegraphics[width=1\linewidth]{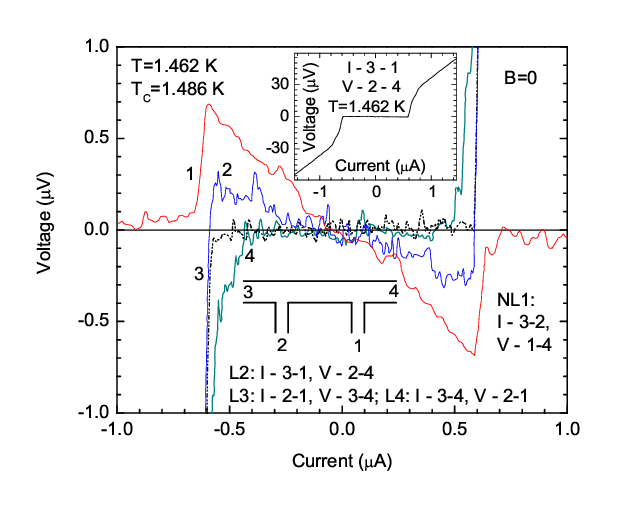}
\caption{\label{f2} (Color online) Line 1 - nonlocal $V_{NL}(I)$
curve recorded according to the measurement circuit ($I$ - 3-2,
$V$ - 1-4). Lines 2-4 - local $V_{L}(I)$ curves recorded according
to different measurement circuits (line 2: $I$ - 3-1, $V$ - 2-4;
3: $I$ - 2-1, $V$ - 3-4; 4: $I$ - 3-4, $V$ - 2-1). Inset at the
top: $V_{L}(I)$ is a curve taken in a larger range of currents and
corresponding to line 2. All curves were taken at $T=1.462$\,K and
$B=0$. Inset at the bottom: a sketch of the structure.}
\end{center}
\end{figure}

We measured the nonlocal $V_{NL}(I)$ (line 1) and local $V_{L}(I)$
(lines 2-4) voltages in the structure biased with direct current
$I$ at different connections of current $I$ and voltage $V$ probes
at $T=1.462$\,K in the zero field (Fig. \ref{f2}). The voltages
$V_{NL}(I)$ and $V_{L}(I)$ were taken from a part of the structure
through which current $I$ did not flow and from another part of
the structure through which current $I$ flowed, respectively. A
sketch of the structure and a measurement lay-out for 1-4 lines -
line 1: $I$ - 3-2, $V$ - 1-4;  2: $I$ - 3-1, $V$ - 2-4;  3: $I$ -
2-1, $V$ - 3-4; 4: $I$ - 3-4, $V$ - 2-1 are shown in the bottom
inset of Fig. \ref{f2}. The upper inset of Fig. \ref{f2}
demonstrates the $V_{L}(I)$ curve ($I$ - 3-1, $V$ - 2-4), measured
in a larger range of currents and corresponding to line 2.

We recorded the $V_{NL}(I)$ and $V_{L}(I)$ curves using an X-Y
coordinate plotter when the applied direct current $I$ swept from
the conditionally positive value $I_{m}$ to the conditionally
negative values $-I_{m}$ and then changed from $-I_{m}$ to
$I_{m}$. The rate of the current sweep is 0.045 $\mu$A/s for all
$V_{NL}(I)$ and $V_{L}(I)$ curves presented in this paper. Figure
\ref{f2} and the top inset show the $V_{NL}(I)$ and $V_{L}(I)$
curves measured when the current changes from $I_{m}$ to $-I_{m}$.
Curves $V_{NL}(I)$ and $V_{L}(I)$ had no hysteresis depending on
the direction of the current sweep. This can be seen from the fact
that the odd function conditions relative to zero $-V_{NL}(-I)
\backsimeq V_{NL}(I)$ and $-V_{L}(-I) \backsimeq V_{L}(I)$ are
held for curves (Fig. \ref{f2}).

To our surprise, the negative nonlocal (line 1) and local (line 2)
voltages appear on the structure when measurement circuits ($I$ -
3-2, $V$ 1-4) and ($I$ - 3-1, $V$ - 2-4) are used, respectively.
With these circuits, a current-carrying part of the structure
includes a wide wire and a narrow wire connected to it, and the
voltage was taken using narrow and wide wires.

Negative nonlocal and local voltages increase approximately
linearly when $I$ increases from zero to critical values of 0.59
and 0.55 $\mu$A, respectively. These voltages have a sign opposite
to the direction of the applied current. With increasing $I$, the
negative nonlocal voltage (line 1) after reaching a peak value of
-0.69 $\mu$V at a current of $I_{c1}=0.59$ $\mu$A, drops sharply
to a value close to zero and remains approximately equal to zero
with a further increase in a current. Whereas, with increasing
$I$, the local voltage (line 2) after reaching a peak value of
-0.31 $\mu$V at a current of $I_{c2}=0.55$ $\mu$A, drops sharply,
changes its sign and demonstrates the usual resistive response
with a resistance close to $R_{0}$, with further increasing $I$
(upper inset of Fig. \ref{f2}).

Lines 3 and 4 show the absence of negative voltage when the
current-carrying part of the structure consists of a wide and two
narrow wires connected to it, and the voltage is recorded using
wide wires ($I$ - 2-1, $V$ - 3-4) and when the current-carrying
part of the structure consists of a wide wire, and the voltage is
recorded using narrow wires ($I$ - 3-4, $V$ - 2-1).

Positive voltage (lines 3 and 4) appears when $I$ reaches critical
values of 0.58 and 0.50 $\mu$A, respectively. With a subsequent
increase in $I$, lines 3 and 4 demonstrate the usual resistive
response (similar to the line on the upper inset of Fig. \ref{f2})
with a resistance close to the normal resistance $R_{0}=37.85$
$\Omega$.

Note that the critical currents of 0.58 (line 3) and 0.50 $\mu$A
(line 4) of the structure are different. A larger value of the
critical current corresponds to the case when the current flows
through a wide and two narrow wires. This is consistent with the
fact that the critical temperature 1.486 K found in the case when
the current $I=0.11$ $\mu$A flows through a wide and two narrow
wires is higher than the critical temperature 1.484 K found in the
case when the current flows through one wide wire. A possible
reason that the critical current (line 3) is higher than the
critical current (line 4) will be suggested in the subsection
\ref{subsection34}.

We believe that the current coordinates of the voltage peaks for
lines 1 and 2, equal to $I_{c1}=0.59$ and $I_{c2}=0.55$ $\mu$A,
respectively, can be identified with the values of the
superconducting critical current. This assumption is confirmed by
the fact that the values of $I_{c1}$ and $I_{c2}$ practically
coincide with the critical current $I_{c3}=0.58$ $\mu$A for line
3.

We calculated the mean nonlocal $R_{NL}$ and local $R_{L}$
negative resistances corresponding to the peak values of the
nonlocal and local voltages by linearly approximating the
approximately linear part of the $V(I)$ curve located between the
voltage maxima and as the ratio $R=dV/dI$ (where $dV$ is the
voltage interval between peaks, $dI$ is the corresponding current
interval). The $R_{NL}$ and $R_{L}$ values determined by the first
method were approximately ten percent less than these values found
by the second method. In this work, $R_{NL}$ and $R_{L}$ are
obtained by the second method. Resistances $R_{NL}$ and $R_{L}$
are equal to -1.16 and -0.56 $\Omega$ for lines 1 and 2,
respectively.

\begin{figure}
\begin{center}
\includegraphics[width=1\linewidth]{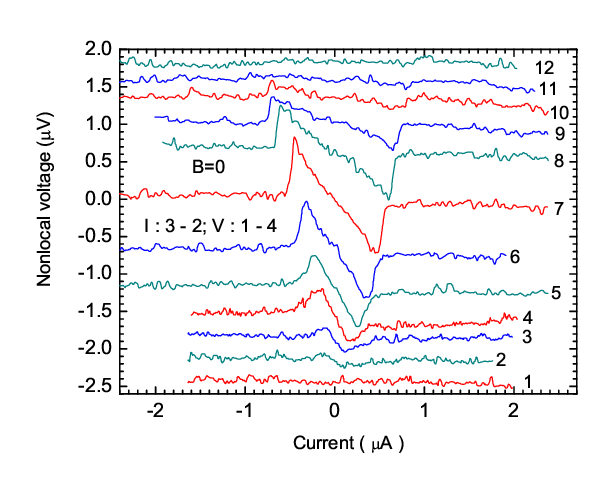}
\caption{\label{f3} (Color online) Lines 1-12 are nonlocal
$V_{NL}(I)$ curves measured according to the measurement circuit
($I$ - 3-2, $V$ - 1-4) in zero field at different $T$,
respectively: 1 - 1.491 K, 2 - 1.486 K, 3 - 1.484 K, 4 - 1.481 K,
5 - 1.476 K, 6 - 1.472 K, 7 - 1.468 K, 8 - 1.462 K, 9 - 1.458 K,
10 - 1.456 K, 11 - 1.453 K, 12 - 1.452 K. Lines 1-6 and 8-12 are
displaced down and up vertically, respectively, relative to line
7.}
\end{center}
\end{figure}

\subsection{Nonlocal $V_{NL}(I)$ voltage at different temperatures $T$ in the zero field}

In order to understand the mechanism and physical parameters of
the appearance of negative voltage (resistance), we recorded both
nonlocal $V_{NL}(I)$ ($I$ - 3-2, $V$ - 1-4) and local $V_{L}(I)$
($I$ - 3-1 , $V$ - 2-4) voltages across the structure at different
$T$ in $B=0$. Curves $V_{NL}(I)$ and $V_{L}(I)$ had almost no
hysteresis when changing the direction of the current sweep. Some
$V_{NL}(I)$ curves (lines 1-12) are shown in Fig. \ref{f3}. Lines
1-6 and 8-12 are displaced down and up vertically, respectively,
relative to line 7. It can be seen that the coordinates of the
voltage peak depend on $T$. $V_{NL}(I)$ is an odd function of $I$.
Moreover, the sign of this voltage is opposite to that of the
applied current. A negative nonlocal voltage (resistance) occurs
at $T<1.491$ K corresponding to the upper part of the resistive
$R_{0}(T)$ transition and disappears at $T=1.452$ K (Figs.
\ref{f3}, \ref{f4} and \ref{f6}).

\subsection{ Nonlocal and local negative resistances as functions of temperature in the zero field}
\label{subsection34}

\begin{figure}
\begin{center}
\includegraphics[width=1\linewidth]{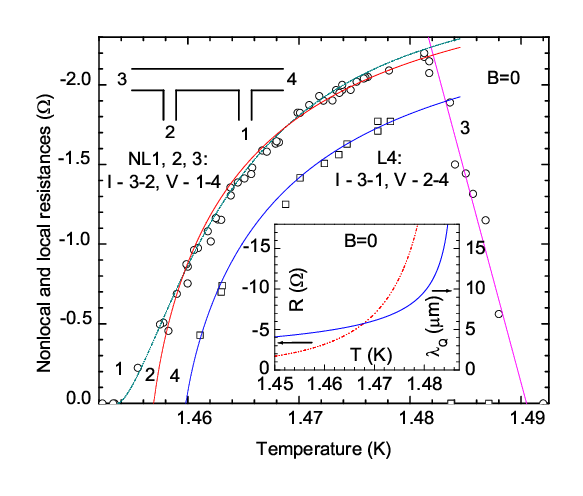}
\caption{\label{f4} (Color online) Measured negative nonlocal
(circles) and local (squares) resistances as functions of
temperature at $B=0$. Lines 1 and 2 are fits of nonlocal
resistance in the range of 1.453 - 1.482 K by the functions
$R_{NL1}(T)$ and $R_{NL2}(T)$ taking into account one
$\sigma^{1AL}$ correction to conductivity and two  $\sigma^{1AL}$
and $\sigma^{1AMT}$ corrections, respectively. Line 3 is a fit of
nonlocal resistance in the range of 1.481 - 1.491 K by the linear
function $R_{NL3}(T)$. Line 4 is a fit of local resistance
(squares) in the range of 1.460 - 1.484 K by the function
$R_{L4}(T)$, which takes into account two $\sigma^{1AL}$ and
$\sigma^{1AMT}$ corrections. Inset below: the dashed and solid
lines show the calculated temperature-dependent negative
resistance $R(x_{0},T)$ and the quasiparticle diffusion length
$\lambda_{Q}(T, 0)$ at $B=0$, respectively. Inset upper: a sketch
of the structure.}
\end{center}
\end{figure}

For the purpose of a detailed analysis of the temperature
dependence of negative nonlocal ($I$ - 3-2, $V$ - 1-4) and local
($I$ - 3-1, $V$ - 2-4) voltages (resistances), we plotted
resistances corresponding to voltage peaks, as well as current and
potential coordinates of nonlocal and local voltage peaks as
functions of $T$ (Figs. \ref{f4}-\ref{f6}).

Figure \ref{f4} shows unexpected experimental nonmonotonic
temperature dependences of negative nonlocal (circles) and local
(squares) resistances corresponding to voltage peaks. Resistances
are calculated from nonlocal $V_{NL}(I)$ and local $V_{L}(I)$
curves measured in several cycles at $B=0$. We found that nonlocal
and local resistances arise in narrow temperature ranges of 1.453
- 1.491 K and 1.460 - 1.484 K, respectively. Lines 1-3 and 4 are
fits for nonlocal and local resistances, respectively. The upper
inset of Fig. \ref{f4} shows a sketch of the structure with the
measurement circuits. The dashed-dotted and solid lines (bottom
inset of Fig. \ref{f4}) represent the theoretical temperature
dependences of the negative resistance $R(x_{0},T)$ and the
quasiparticle diffusion length $\lambda_{Q}(T,0)$ at $B=0$,
respectively. These dependences were obtained in the framework of
the nonequilibrium model proposed by W. Skocpol, M. Beasley, and
M. Tinkham (SBT) \cite{sbt}.

In order to explain the measured negative nonlocal (line 1 in Fig.
\ref{f2}) and local (line 2 in Fig. \ref{f2}) voltages
(resistances) and temperature dependences of negative nonlocal
(circles in Fig. \ref{f4}) and local (squares in Fig. \ref{f4})
resistances, we suggested that the critical temperatures $T_{cn}$
and $T_{cw}$ of the narrow and wide wires, respectively, forming
our structure, are different. In addition, we believe that these
critical temperatures are close to 1.453 and 1.491 K (Fig.
\ref{f4}), which determine the temperature range for the existence
of negative nonlocal and local resistances. Then, in the range of
1.453 - 1.491 K, our structure, taken together with current and
potential wires is a heterogeneous N-S structure. Thus, negative
nonlocal (line 1 in Fig. \ref{f2}, Fig. \ref{f3}, circles in Fig.
\ref{f4}) and local (line 2 in Fig. \ref{f2}, squares in Fig.
\ref{f4}) resistances are observed only in cases when the
current-carrying part of the structure consists of a
superconducting and normal wires, and the voltage is recorded
using normal and superconducting wires.

In order to verify the supposition about different critical
temperatures $T_{cn}$ and $T_{cw}$ of narrow and wide wires, the
critical temperature of the reference narrow wire with a width of
$w_{n}=0.27$ $\mu$m fabricated at the same time with the structure
under study was measured. The critical temperature of the
reference narrow wire, found in the middle of the resistive $R(T)$
transition (not shown here) of the wire part with the length of
6.69 $\mu$m at 0.11 $\mu$A current, is 1.455 K. The value of 1.455
K is in agreement with the assumed values of the critical
temperature equal to 1.453 K and 1.460 K (Fig. \ref{f4}), which
can be attributed to the critical temperatures $T_{cn2}$ and
$T_{cn1}$ for narrow wires 2 and 1 of the structure, respectively.
Note that the critical temperatures of the structure under study,
found in the middle of the $R_{0}(T)$ transition and recorded in
accordance with two measurement circuits: $I$ - 3-1, $V$ - 2-4,
and $I$ - 3-4, $V$ - 2-1 are equal to 1.486 and 1.484 K,
respectively. The second measurement circuit corresponds to the
case when the current-carrying part of the structure is a wide
wire, and the voltage is recorded using narrow wires. Then, the
critical value of 1.491 K (Fig. \ref{f4}) can be equated to the
critical temperature $T_{cw}$ of the wide wire. Thus, we
unexpectedly found that $T_{cn}<T_{cw}$. The above inequality was
confirmed by measuring the critical temperatures of other aluminum
narrow and wide wires with a thickness of $d \backsimeq19$ nm and
widths of $w_{n} \backsimeq 0.27$ and $w_{w} \backsimeq 0.5$ $
\mu$m. The critical temperatures of these narrow and wide wires
were equal to 1.447 and 1.490 K, respectively.

Thus, in the temperature range $T_{cn}<T<T_{cw}$, negative
nonlocal and local voltages (resistances) appear at $I\leqslant
I_{c}$ when the voltage is measured by a narrow normal and wide
superconducting wires, and the current-carrying part of the
structure includes both wide superconducting wire (where
nonequilibrium fluctuations of the superconducting order parameter
appear at $T<T_{cw}$ \cite{sbt}) and a narrow normal wire (where
equilibrium superconducting fluctuations occur at $T>T_{cn}$
\cite{larkin}). Therefore, we applied both the nonequilibrium SBT
model \cite{sbt} and the equilibrium model \cite{larkin} to
theoretically describe the temperature dependences of negative
nonlocal and local resistances.

In the range of 1.455 - 1.491 K, for any measurement circuits, the
critical current of our N-S structure is nonlocal and depends on
the region located on both sides of the points of connection of
the narrow wire to the wide wire. In the case when the applied
current passes through a narrow and wide wires, the measured
critical current is determined both by the part of the narrow
wire, where superconductivity is induced due to the proximity
effect, and by the part of the wide wire, where superconductivity
is suppressed.

Let us speculate why the critical current (line 3 in Fig.
\ref{f2}) recorded according to the measurement circuit ($I$ -
2-1, $V$ - 3-4) was unexpectedly larger than the critical current
(line 4 in Fig. \ref{f2}) recorded according to a different
measurement circuit ($I$ - 3-4, $V$ - 2-1). The critical current
of the N-S structure depends on the density of superconducting
pairs $n_{s}$ near the points of connection of the narrow normal
wire to the wide superconducting wire. An increase in the applied
current flowing through a wide wire and two narrow wires attached
to it leads to a decrease in $n_{s}$. However, due to the
proximity effect, superconducting pairs come from the non-current
carrying ends of the wide wire, to the points of connection of the
narrow wire to the wide wire, and slightly compensate for the
decrease in $n_{s}$. Such compensation for the decrease in $n_{s}$
is either absent or much less in the case when the current flows
only through a wide wire, and the voltage is measured using narrow
wires ($I$ - 3-4, $V$ - 2-1). Thus, it should be expected that the
critical temperature and current of the structure will be higher
in the case ($I$ - 2-1, $V$ - 3-4) than the critical temperature
and current in the other case ($I$ - 3-4, $V$ - 2-1).

Further, within the framework of the nonequilibrium SBT model, we
suggest a mechanism for the appearance of negative voltages
(resistances) in the structure at $T<T_{c}=1.486$ K and
$I\leqslant I_{c}$. In addition, we calculate the temperature and
magnetic-field dependences of negative voltages (resistances). In
the interval $T_{cn}<T<T_{cw}$, N-S interfaces are formed at the
points of connection of narrow normal wires (2 and 1) to a wide
superconducting wire (3-4). The applied current $I$ flowing
through one of the N-S interfaces, creates nonequilibrium regions
in the superconductor with a nonlocal or local quasiparticle
charge imbalance. The quasiparticle charge imbalance appears due
to the injection of quasiparticles from a narrow current wire with
a lower critical temperature ($T_{cn2}=1.453$ and $T_{cn1}=1.460$
K for wires 2 and 1, respectively) into a wide wire (3-4) with a
higher critical temperature $T_{cw}=1.491$ K.

A local quasiparticle charge imbalance appears in that part of the
superconductor through which the applied current $I$ flows. The
current $I$ flowing in the region of the quasiparticle charge
imbalance created by the phase-slip center in a
quasi-one-dimensional superconducting structure at $I\geqslant
I_{c}$ consists of the sum of the spatially dependent
time-averaged normal (quasiparticle) $\bar{I}_{N}(x)$ and
superconducting $\bar{I}_{S}(x)$ currents \cite{sbt}. Where $x$ is
the longitudinal coordinate measured from the core of the
phase-slip center  in the conditionally positive and negative
directions. The time-averaged superconducting current in the  PSC
core is $\bar{I}_{S}(x=0) \backsimeq 0.5I_{c}$ at $I=I_{c}$
\cite{sbt}. However, for both the local and nonlocal cases in our
hybrid N-S structure, the time-averaged superconducting current at
the center of the N-S interface $\bar{I}_{S}(x=0)$ is close to
zero at $I<I_{c}$.

A nonlocal quasiparticle charge imbalance appears in another part
of the superconductor through which the applied current is zero.
Nevertheless, the nonlocal quasiparticle current and the
superconducting countercurrent, which balance each other
($I=\bar{I}_{N}(x)-\bar{I}_{S}(x)=0$), flow in this part of the
superconductor.

With increasing $I$, both local and nonlocal quasiparticle
currents reach a critical value equal to the critical
superconducting current, and then sharply decrease. This critical
value corresponds to the peaks of negative nonlocal (line 1 in
Fig. \ref{f2}) and local (line 2 in Fig. \ref{f2}) voltages.

We believe that there is some difference between negative nonlocal
and local voltages in the current interval between voltage peaks
(lines 1 and 2 in Fig. \ref{f2}) due to a small difference in the
critical temperatures of narrow wires 2 and 1.

Nonequilibrium phenomena in which superconducting and
quasiparticle currents coexist in quasi-one-dimensional
superconducting structures \cite{sbt} and structures with a N-S
interface \cite{yu} are described using the electrochemical
potentials of superconducting pairs $\mu_{p}$ and quasiparticles
$\mu_{q}$. For the first time, these potentials were postulated
and used to study phase-slip centers \cite{sbt}.

The time-averaged electrochemical potentials of superconducting
pairs $\bar{\mu}_{p}$ and quasiparticles $\bar{\mu}_{q}$ are
measured using the superconducting and normal-metal probes,
respectively. The potentials $\bar{\mu}_{p}$ and $\bar{\mu}_{q}$
differ strongly in the nonequilibrium region and coincide far
beyond the nonequilibrium region. In \cite{sbt}, it was assumed
that these spatially dependent potentials are described by the
expressions
$\bar{\mu}_{p}(I,x,T)=e\lambda_{Q}(T)\rho_{n}A^{-1}\bar{I}_{N}(x=0)tanh(x/\xi(T))$
and
$\bar{\mu}_{q}(I,x,T)=e\lambda_{Q}(T)\rho_{n}A^{-1}\bar{I}_{N}(x=0)tanh(x/\lambda_{Q}(T))$
(here the $x$ coordinate axis has a beginning in a core of the
phase-slip center, $A$ is the cross-sectional area of the
microstructure, $\bar{I}_{N}(x=0)$ is the time-averaged normal
(quasiparticle) current at the center of the N-S interface). A
significant change in the potentials $\bar{\mu}_{p}$ and
$\bar{\mu}_{q}$ occurs at the distances $\xi(T)$ and
$\lambda_{Q}(T)$ from the center of the nonequilibrium region,
respectively. The experiment \cite{dolan} confirmed this
assumption.

Nonequilibrium regions with a quasiparticle charge imbalance are
formed in our structure when the applied direct current $I<I_{c}$
flows through one (or two) N-S interfaces of the structure with
three measurement circuits (lines 1-3 in Fig. \ref{f2}). However,
in order to observe a nonzero voltage due to the quasiparticle
charge imbalance at currents satisfying the requirement
$0<I\leqslant I_{c}$, it is necessary that the potential pair
consisted of normal and superconducting wires. This situation is
realized in the cases of lines 1 and 2 in Fig. \ref{f2}. In cases
(lines 3 and 4 in Fig. \ref{f2}), the usual positive voltage is
recorded at currents $I\geqslant I_{c}$.

We believe that the negative nonlocal (or local) voltage measured
on our structure corresponds to the difference between the
electrochemical potentials of quasiparticles and superconducting
pairs
$V(I,x,T)=(\bar{\mu}_{q}(I,x,T)-\bar{\mu}_{p}(I,x,T))/e=\lambda_{Q}(T)\rho_{n}A^{-1}\bar{I}_{N}(x=0)
(tanh(x/\lambda_{Q}(T))-tanh(x/\xi(T)))$, where $\bar{I}_{N}(x=0)
\backsimeq I$ for nonlocal and local (at $I\leqslant I_{c}$)
cases. The $x$ axis is directed from the center of the N-S
interface into the interior of the superconductor. Then, at a
distance $x=x_{0}>\xi(T)$ from the center of the N-S interface
(here $x_{0}=6.69$ $\mu$m for both nonlocal and local cases),
$V(I,x_{0},T)=-\lambda_{Q}(T)\rho_{n}A^{-1}I(1-tanh(x_{0}/\lambda_{Q}(T)))$.
Voltage $V(I,x_{0},T)<0$, since
$\bar{\mu}_{q}(I,x_{0},T)<\bar{\mu}_{p}(I,x_{0},T)$. If the
spatial variation of the quasiparticle charge imbalance in
nonlocal and local cases is the same, then it should be expected
that the curves $V_{NL}(I)$ and $V_{L}(I)$ will be coincide at
currents $I\leqslant I_{c}$.

The expression for the temperature dependence of the negative
nonlocal and local resistances
$R(x_{0},T)=-\lambda_{Q}(T)\rho_{n}A^{-1}(1-tanh(x_{0}/\lambda_{Q}(T)))$
follows from the expression for $V(I,x_{0},T)$. For calculations,
it is necessary to know the temperature-dependent length of
quasiparticle diffusion in the zero field $\lambda_{Q}(T,0)$.

Next, we calculated the quasiparticle diffusion length
$\lambda_{Q}(T, B)$ as a function of the temperature $T$ and the
field $B$ using the expression $\lambda_{Q}(T, B)=
\sqrt{D\tau_{Q}}$, where $\tau_{Q}
=4kT\sqrt{\tau_{e-p}/2\Gamma}/\pi\Delta(T,B)$ is the relaxation
time of the quasiparticle charge imbalance \cite{stuivinga},
$\Delta(T,B)$ is the superconducting gap as a function of $T$ and
$B$, $\tau_{e-p}$ is the time of inelastic electron-phonon
collisions,
$\Gamma=(2\tau_{e-p})^{-1}+\tau_{jc}^{-1}+\tau_{s}^{-1}$,
$\tau_{jc}^{-1}(T,B)=D/6\xi_{GL}^{2}(T,B)=(1-(B^{2}/B_{c}^{2}(T))/6\tau_{GL}$
is the relaxation frequency of the quasiparticle charge imbalance
due to the superconducting current \cite{stuivinga},
$\tau_{GL}=\pi\hbar/(8k(T_{c}(0)-T))$ is the Ginzburg-Landau time,
$\tau_{s}^{-1}(T,B)=
(\Delta(0,0)/\hbar)(B^{2}/B_{c}^{2}(0))=(B^{2}/B_{c}^{2}(T))/\tau_{GL}$
is the depairing frequency of superconducting pairs by a magnetic
field  \cite{kadin}. Near the critical temperature, the gap
\cite{tinkham} is equal to
$\Delta(T,B)=\Delta(T,0)(1-B^{2}/B_{c}^{2}(T))^{1/2}$, where
$\Delta(T,0)=1.74\sqrt{2/3}\Delta(0,0)(1-T/T_{c})^{1/2}$ at the
critical current, $\Delta(0,0)=1.764kT_{c}$. Here, $B_{c}(0)$ and
$B_{c}(T)=\sqrt{3}\Phi_{0}/\pi w\xi_{GL}(T)$ are the maximum
critical fields for a quasi-one-dimensional superconducting wire
with a width $w$ in a magnetic field perpendicular to the
substrate surface at $T=0$ and at a given temperature $T$,
respectively.

We plotted the length $\lambda_{Q}(T, 0)$  as a function of $T$ in
the zero field (lower inset of Fig. \ref{f4}) and the length
$\lambda_{Q}(T_{2},B)$ as a function of the field $B$ at
$T_{2}=1.468 $\,K (lower inset of Fig. \ref{f10}), taking into
account $\tau_{jc}^{-1}$, since $\tau_{jc}^{-1}>>\tau_{e-p}^{-1}$
for aluminum in our case. $\lambda_{Q}(T_{2},0)=5.75$ $\mu$m  from
these dependences.

The calculated temperature dependences of the negative resistance
$R(x_{0},T)$ at $B=0$ are shown in the lower inset of Fig.
\ref{f4}. It is seen that at temperatures of 1.45-1.47 K, the
negative resistance $R(x_{0},T)$ and the measured negative
nonlocal and local resistances have the same sign and are close in
order of a magnitude. However, in the whole range of 1.453-1.491
K, the experimental temperature dependences of the resistances
radically differ from the theoretical dependence $R(x_{0},T)$
(Fig. \ref{f4} and the lower inset). The main reason for this
difference may be that we have not taken into account the
equilibrium superconducting fluctuations at $T>T_{cn}=1.453
(1.460)$\,K  in normal wires \cite{larkin} of the current-carrying
part of the structure.

Note that the measured negative resistances in the range of
1.453-1.484 K (Fig. \ref{f4}) show temperature dependences similar
(except for the negative sign of resistances) to the temperature
dependence of the resistive transition $R_{0}(T)$ of the structure
(Fig. \ref{f1}) described by the theory of equilibrium
superconducting fluctuations at $T>T_{c}$ \cite{larkin}.

We believe that equilibrium superconducting fluctuations at
$T>T_{cn}=1.453$\,K occurring in a narrow normal current wire
affect the electron transport in a wide superconducting current
wire. Therefore, we use the equilibrium model \cite{larkin} to fit
the measured negative nonlocal resistance (circles in Fig.
\ref{f4}) in the temperature range of 1.453 - 1.482 K by two
functions $R_{NL1}(T)$ and $R_{NL2}(T)$.

Dash-dotted line 1 is the first fitting function
$R_{NL1}(T)=R_{f1}/(1+C_{2}\rho_{n}\sigma^{1AL}(0)(T/T_{cf1}-1)^{-3/2})$,
which takes into account only the $\sigma^{1AL}$ correction
\cite{larkin} at $T>1.453 $\,K. The fitting resistance is
$R_{f1}=-2.8$ $\Omega$, the fitting critical temperature is
$T_{cf1}=1.454$\,K, and the multiplier $C_{2}=21.8$, which
characterizes the deviation of the experimental correction to the
conductivity from the $\sigma^{1AL}(0)$ correction.

Solid line 2 is the second fitting function
$R_{NL2}(T)=R_{f2}/(1+\rho_{n}(\sigma^{1AL}+\sigma^{1AMT}))=R_{f2}/(1+\rho_{n}\sigma^{1AL}(0)
\tau^{-3/2}(1+8(\tau/\gamma_{f2})/(1+\sqrt{\tau/\gamma_{f2}})))$
(where $\tau=T/T_{cf2}-1$), taking into account two $\sigma^{1AL}$
and $\sigma^{1AMT}$ corrections \cite{larkin}. We found fitting
values $\gamma_{f2}=0.001$, $R_{f2} =-3.0$ $\Omega$, and
$T_{cf2}=1.457$\,K. Taking the total phase-breaking frequency of
the electron wave function $\tau_{\varphi}^{-1}=\tau_{N}^{-1}+
\tau_{e-p}^{-1}=7.9\times10^{8}$ $s^{-1}$ at $T=1.457$ K, we
obtained the calculated depairing factor $\gamma=0.0016$ for the
wire $w_{w}=0.5$ $\mu$m. The factor $\gamma$ is 1.6 times greater
than the factor $\gamma_{f2}$.

The negative nonlocal resistance (circles) in the range of 1.481 -
1.491 K is fitted by the linear function
$R_{NL3}(T)=R_{f3}(0)(1-T/T_{cf3})$, where the fitting
$R_{f3}(0)=-388$ $\Omega$ and $T_{cf3}=1.491 $\,K  (line 3 in Fig.
\ref{f4}). The nonlocal resistance sharply tends to zero as $T$
approaches to $T_{cw}$.

Within the framework of the theory of superconducting fluctuations
above $T_{cn}=1.460$ K, we fitted the negative local resistance
(squares) in the range of 1.460 - 1.484 K by the function
$R_{L4}(T)=R_{f4}/(1+\rho_{n}\sigma^{1AL}(0)\tau^{-3/2}(1+8(\tau/\gamma_{f4})/(1+\sqrt{\tau/\gamma_{f4}})))$
(where $\tau=T/T_{cf4}-1$), taking into account two $\sigma^{1AL}$
and $\sigma^{1AMT}$ corrections \cite{larkin}. Fitting  values
$\gamma_{f4}=0.0008$, $R_{f4}=-2.7$ $\Omega$, and
$T_{cf4}=1.460$\,K. The calculated depairing factor
$\gamma=0.0016$ is twice as large as $\gamma_{f4}$. Line 4 in Fig.
\ref{f4} represents the function $R_{L4}(T)$. Using the original
expression $\sigma^{1AMT}$ \cite{thompson} to plot $R_{NL2}(T)$
and $R_{L4}(T)$ functions results in worse approximations (not
shown here).

Thus, taking into account the equilibrium superconducting
fluctuations above $T_{cn}$ \cite{larkin}  allows to qualitatively
describe the measured negative nonlocal and local resistances
(except for the negative sign of the resistances) in the range of
1.453 (1.460) - 1.484 K (Fig. \ref{f4}).

So, in the interval $T_{cn}<T<T_{cw}$ (with the exception of $T$
that are very close to $T_{cw}$), both nonequilibrium \cite{sbt}
and equilibrium models \cite{larkin} qualitatively describe the
temperature dependences of negative nonlocal and local
resistances. However, for a more adequate understanding, it is
necessary to create a model that combines nonequilibrium and
equilibrium models. In addition, it is necessary to take into
account the possible influence of the normal and a part of the
superconducting wires that compose the potential pair on the
temperature dependence of the resistance.

\subsection{Nonlocal and local critical currents, maximum negative voltages as functions of $T$ in the zero field}

\begin{figure}
\begin{center}
\includegraphics[width=1\linewidth]{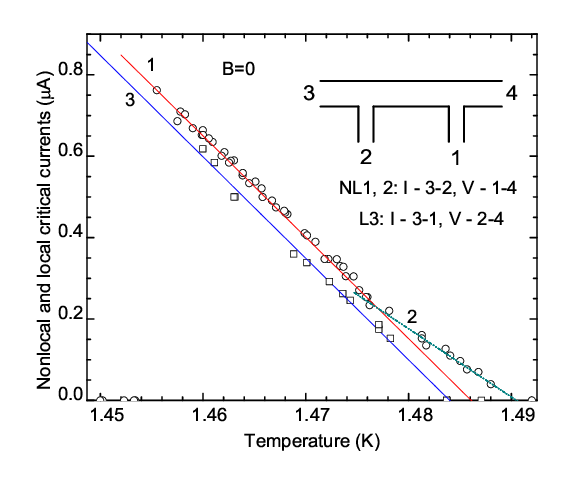}
\caption{\label{f5} (Color online) Measured nonlocal (circles) and
local (squares) critical currents as functions of $T$ in $B=0$.
Lines 1 and 2 are linear fits of the nonlocal critical current
(circles) in two ranges of $1.455 -1.477$\,K and $1.475 -1.491$\,K
by the functions $I_{cNL1}(T)$ and $I_{cNL2}(T)$, respectively.
Line 3 is a linear fit of the local critical current (squares) by
the $I_{cL3}(T)$ function in the range of $1.460 - 1.484$\,K.
Inset: a sketch of the structure.}
\end{center}
\end{figure}

\begin{figure}
\begin{center}
\includegraphics[width=1\linewidth]{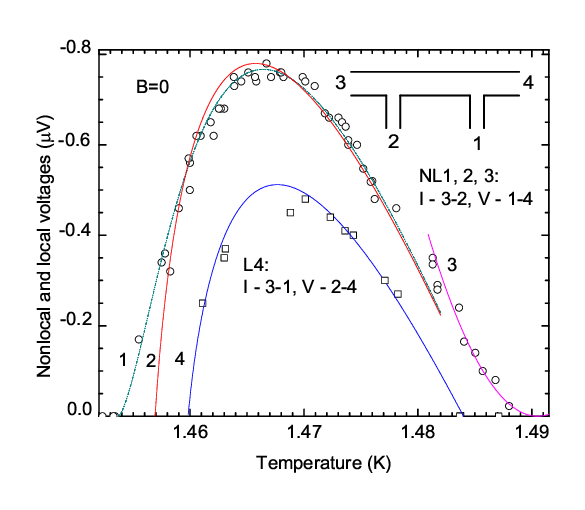}
\caption{\label{f6} (Color online) Measured maximum (peak) values
of negative nonlocal (circles) and local (squares) voltages as
functions of $T$ in $B=0$. Lines 1 and 2 are the fits of the
nonlocal voltage (circles) in the range of $1.453 - 1.478$\,K by
two functions $V_{NL1}(T)$ and $V_{NL2}(T)$, taking into account
$\sigma^{1AL}$ correction to the conductivity  and two
$\sigma^{1AL}$ and $\sigma^{1AMT}$ corrections, respectively. Line
3 is the fit of the other part of the nonlocal voltage (circles)
in the range of $1.481 - 1.491$\,K by the function $V_{NL3}(T)=
R_{NL3}(T)I_{cNL2}(T)$. Line 4 is the fit of the local resistance
(squares) in the range of $1.460 - 1.484$\,K by the function
$V_{L4}(T)$, taking into account both $\sigma^{1AL}$ and
$\sigma^{1AMT}$ corrections. Top inset: sketch of the structure.}
\end{center}
\end{figure}

The experimental and theoretical temperature dependences of the
resistances corresponding to the peak values of nonlocal and local
voltages have been presented previously (Fig. \ref{f4}). Here we
continue the series of Figs. \ref{f4}-\ref{f6}.

Figure \ref{f5} shows the measured nonlocal (circles) and local
(squares) critical currents, which we identify with the current
coordinates of the peaks of nonlocal and local voltages as
functions of the temperature at $B=0$. The data are obtained from
nonlocal $V_{NL}(I)$  ($I$ - 3-2, $V$ - 1-4) and local $V_{L}(I)$
($I$ - 3-1, $V$ - 2-4) curves, recorded in several cycles in the
zero field. The inset of Fig. \ref{f5} demonstrates a sketch of
the structure.

Nonlocal and local critical currents are approximated by linear
functions (lines 1-3). Solid line 1 is the fit of the nonlocal
critical current (circles) in the range of $1.455 - 1.477$\,K by
the function $I_{cNL1}(T)=I_{cf1}(0)(1-T/T_{cNLf1})$ (where
$I_{cf1}(0)=37$\,$\mu$A and $T_{cNLf1}=1.486$\,K are the fitting
critical current at $T=0$\,K and the critical temperature,
respectively). Dash-dotted line 2 is the fit of the nonlocal
critical current (circles) in another range of $1.475 - 1.491$\,K
by the function $I_{cNL2}(T)=I_{cf2}(0)(1-T/T_{cNLf2})$ (where
$I_{cf2}(0)=25$ $\mu$A  and $T_{cNLf2}=1.491$\,K). Solid line 3 is
the fit of the local critical current (squares) in the range of
$1.460 - 1.484$\,K by the function
$I_{cL3}(T)=I_{cf3}(0)(1-T/T_{cLf3})$ (where $I_{cf3}(0)=37$
$\mu$A , $T_{cLf3}=1.484$\,K). It can be seen that the
$I_{cL3}(T)$ data are lower than the $I_{cNL1}(T)$ data.

Let us note that the fitting functions $R_{NL3}(T)$ and
$I_{cNL2}(T)$ for the nonlocal resistance in the range of
$1.481-1.491$\,K (line 3 in Fig. \ref{f4}) and the nonlocal
critical current in the range of $1.475-1.491$\,K (line 2 in Fig.
\ref{f5}), respectively, have the same linear temperature
dependence. Using the expressions for $R_{NL3}(T)$ and
$I_{cNL2}(T)$, we found that $R_{NL3}(T)=k_{3}I_{cNL2}(T)$ (here
$k_{3}=-15.5$\,$\Omega$/$\mu$A is an fitting constant). Thus, in
the range of $1.475-1.491$\,K, the nonlocal resistance is directly
proportional to the nonlocal critical current.

In addition to linear fits of the temperature dependences of the
nonlocal and local critical currents, we used nonlinear fits in
the framework of the GL theory by the functions $I_{cNL4}(T)$ and
$I_{cL5}(T)$ (not shown in Fig. \ref{f5}).

The function $I_{cNL4}(T)=I_{cf4}(0)(1-T/T_{cNLf4})^{3/2}$ (where
fitting $I_{cf4}(0)=180$ $\mu$A and $T_{cNLf4}=1.495$\,K) does not
approximate well enough the nonlocal critical current (circles) in
the range of 1.455 - 1.491 K. The fit of the local critical
current (squares) by the function
$I_{cL5}(T)=I_{cf5}(0)(1-T/T_{cLf5})^{3/2}$, (where
$I_{cf5}(0)=180$ $\mu$A, $T_{cLf5}=1.493$\,K) is also worse than
the linear fit. The fitting $T_{cNLf4}=1.495$ and
$T_{cLf5}=1.493$\,K are higher than $T_{c}=1.486$\,K, determined
in the middle of the resistive N-S transition. Moreover, the
fitting critical current $I_{cf4}(0)=I_{cf5}(0)=180$ $\mu$A is
much lower than the GL depairing critical current $I_{GL}(0)=300$
$\mu$A for a narrow wire $w_{n}=0.27$ $\mu$m wide and $d=19$ nm
thick. Note that the critical current (Fig. \ref{f5}) was measured
in our structure, the current-carrying part of which consists of a
narrow normal and wide superconducting wires. Therefore, the
critical current is determined both by the section of the narrow
wire where the superconducting order parameter is induced due to
the proximity effect, and by the section of the wide
superconducting wire where the order parameter is suppressed.

Thus, the temperature dependences of the nonlocal and local
critical currents are not well described by the GL theory in the
range of 1.455 - 1.491 K.

Additional measurements (not shown here) of the local $V(I)$
curves ($I$ - 3-1, $V$ - 2-4) of the structure at $T<1.4$\,K (when
negative nonlocal and local voltages are not observed) showed that
the critical current is well fitted within the framework of the GL
theory by the function $I_{GL}(T)=I_{cf}(0)(1-T/T_{cf})^{3/2}$
(where the fitting critical current $I_{cf}(0)=330$ $\mu$A is
close to $I_{GL}(0)$, $T_{cf}=1.435$\,K is probably equal to the
critical temperature $T_{cn}$ of a single narrow wire without the
influence of the wide wire).

Let us speculate about the reasons for the deviation of the
temperature dependence of the measured critical current (Fig.
\ref{f5}) from the dependence
$I_{GL}(T)=I_{GL}(0)(1-T/T_{c})^{3/2}$ in the range of 1.455 -
1.491 K. A possible reason for low critical current
($I_{c}(0)=180$ $\mu$A) can be that the nonequilibrium region of
the structure with a low superconducting order parameter $\Delta$
is in a more dissipative state \cite{kuznjetplet19} due to poor
heat removal of overheated quasiparticles from this region.

Another reason for the low critical current and linear temperature
dependence may be the appearance of the Josephson junction with a
length equal to the width of a narrow normal wire 0.27 $\mu$m, at
the region of connection of a narrow normal wire to a wide
superconducting wire.

The critical current of the Josephson junction with a weak link
(short microconstriction) and identical superconducting banks near
$T_{c}$ is determined by the expression \cite{likharev}
$I_{J}(T)=\pi\Delta^{2}(T)/4ekTR_{J}=I_{J}(0)(1-T/T_{c})$, where
$R_{J}$ is the resistance of the Josephson junction. We believe
that the fitting values of the critical current at $T=0$\,K and
critical temperature: $I_{cf1}(0)=37$\,$\mu$A and
$T_{cNLf1}=1.486$\,K (line 1 in Fig. \ref{f5}), $I_{cf2}(0)=25$
$\mu$A and $T_{cNLf2}=1.491$\,K (line 2 in Fig. \ref{f5}),
$I_{cf3}(0)=37$ $\mu$A and $T_{cLf3}=1.484$\,K (line 3 in Fig.
\ref{f5}) can be identified with the corresponding values of
$I_{J}(0)$ and $T_{c}$ in the expression for $I_{J}(T)$. Critical
currents 37 $\mu$A and 25 $\mu$A correspond to the Josephson
resistances equal to $R_{J1,3}=0.676R_{0}=25.6$ $\Omega$ and
$R_{J2}=R_{0}=38.0$ $\Omega$, respectively. These resistances
correspond to quite adequate values of the wire lengths 4.57
$\mu$m and $L_{21}=6.69$ $\mu$m, respectively. Despite the fact
that the central region of the Josephson junction, where $\Delta$
is substantially reduced, has a length shorter than $\xi(T)$, the
resistance $R_{J}$  is determined by the large quasiparticle
diffusion length $\lambda_{Q}(T)$.

Thus, the linear temperature dependences of the critical current
can be explained by the formation of the Josephson junction in our
structure.

Figure \ref{f6}, which completes the series of Figs.
\ref{f4}-\ref{f6}, shows the non-monotonic temperature dependences
of the maximum negative nonlocal (circles) and local (squares)
voltages corresponding to the peak voltage values. Data are taken
from nonlocal $V_{NL} (I)$ and local $V_{L} (I)$ curves recorded
in several cycles in the zero field. Lines 1-3 are fitting
functions for nonlocal voltage. Line 4 is an fitting function for
local voltage. The inset of Fig. \ref{f6} demonstrates a sketch of
the structure.

Fitting functions for negative nonlocal
$V_{NL}(T)=R_{NL}(T)I_{cNL}(T)$ and local
$V_{L}(T)=R_{L}(T)I_{cL}(T)$ voltages are obtained by multiplying
the corresponding fitting functions for nonlocal and local
resistances (Fig. \ref{f4}) and critical currents (Fig. \ref{f5}).

Fitting $V_{NL}(T)$ functions for two temperature ranges of $1.453
- 1.478$\,K and $1.481 - 1.491$\,K are fundamentally different
from each other.

In the range of $1.453 - 1.478$\,K, we fitted the nonlocal voltage
(circles) by two functions $V_{NL1}(T)=R_{NL1}(T)I_{cNL1}(T)$  and
$V_{NL2}(T)=R_{NL2}(T)I_{cNL1}(T)$, taking into account
$\sigma^{1AL}$ correction to the conductivity  and two
$\sigma^{1AL}$ and $\sigma^{1AMT}$ corrections, respectively. The
expression $I_{cNL1}(T)=I_{cf1}(0)(1-T/T_{cNLf1})$ (where
$I_{cf1}(0)=37$ $\mu$A and $T_{cNLf1}=1.486$\,K), was taken as an
fitting function for the nonlocal critical current (Fig.
\ref{f5}).

In the range of $1.481 - 1.491$\,K, we approximated the nonlocal
voltage (circles) with the function
$V_{NL3}(T)=R_{NL3}(T)I_{cNL2}(T)$. The function
$R_{NL3}(T)=k_{3}I_{cNL2}(T)$, where
$k_{3}=-15.5$\,$\Omega$/$\mu$A and the function
$I_{cNL2}(T)=I_{cf2}(0)(1-T/T_{cNLf2})$, where $I_{cf2}(0)=25$
$\mu$A and $T_{cNLf2}=1.491$\,K, are shown in Figs. \ref{f4} and
\ref{f5}, respectively.

\begin{figure}
\begin{center}
\includegraphics[width=1\linewidth]{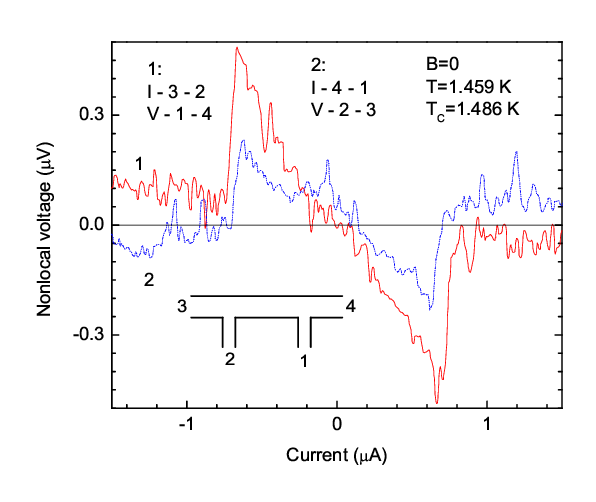}
\caption{\label{f7} (Color online) Nonlocal voltages $V_{NL1}(I)$
(solid line 1) and $V_{NL2}(I)$ (dash-dotted line 2) as functions
of $I$ at $T=1.459$\,K in the zero field, recorded in the right
and left parts of the structure with two measurement circuits ($I$
- 3 - 2, $V$ - 1 - 4) and ($I$ - 4 - 1, $V$ - 2 - 3),
respectively. Inset: a sketch of the structure.}
\end{center}
\end{figure}

In the range of $1.460 - 1.484$\,K, the local voltage (squares)
was fitted by the function $V_{L4}(T)=R_{L4}(T)I_{cL3}(T)$, that
is, by the multiplication of the fitting function for resistance
$R_{L4}(T)$, which takes into account both $\sigma^{1AL}$ and
$\sigma^{1AMT}$ corrections (Fig. \ref{f4}), and the fitting
function for the local critical current $I_{cL3}(T)=
I_{cf3}(0)(1-T/T_{cLf3})$, where $I_{cf3}(0)=37$ $\mu$A and
$T_{cLf3}=1.484$\,K (Fig. \ref{f5}). It is seen that the function
$V_{L4}(T)$ lies lower than $V_{NL2}(T)$.
\subsection{Checking the geometric symmetry of the nonlocal effect}

The geometric symmetry of the nonlocal effect was checked by
recording the nonlocal $V_{NL1}(I)$ (solid line 1) and
$V_{NL2}(I)$ (dash - dotted line 2) curves from the right and left
parts of the structure at $T=1.459$\,K in the zero field with two
measurement circuits ($I$ - 3 - 2, $V$ - 1 - 4) and ($I$ - 4 - 1,
$V$ - 2 - 3), respectively (Fig. \ref{f7} and inset).

Usually, the nonlocal voltage was detected according to the
measurement circuit ($I$ -3 - 2, $V$ -1 - 4). It can be seen that
the $V_{NL1}(I)$  and $V_{NL2}(I)$ curves are slightly different
and have different coordinates of voltage peaks ($I_{c1}=0.66$
$\mu$A, $V_{1}=-0.49$ $\mu$V) and ($I_{c2}=0.62$ $\mu$A,
$V_{2}=-0.23$ $\mu$V), respectively. The resistances corresponding
to the voltage peaks are $R_{1}=-0.73$ $\Omega$  and $R_{2}=-0.38$
$\Omega$ for lines 1 and 2, respectively.

We believe that the asymmetry of the electron transport in the
wire occurs due to the different critical temperatures $T_{cn1}$
and $T_{cn2}$ ($T_{cn1}<T_{cn2}$) of narrow wires 1 and 2,
respectively. For the same reason, the local $V_{L}(I)$ curve
(line 2 in Fig. \ref{f2}), taken according to the measurement
circuit ($I$ -3-1, $V$ -2-4), has lower values of current, voltage
and resistance corresponding to the peak voltage than the nonlocal
$V_{NL1}(I)$ ($I$ - 3-2, $V$ - 1-4) curve (line 1 in Fig.
\ref{f2}).

\subsection{Nonlocal $V_{NL}(I)$ curves in different magnetic fields at $T=1.468$\,K}

\begin{figure}
\begin{center}
\includegraphics[width=1\linewidth]{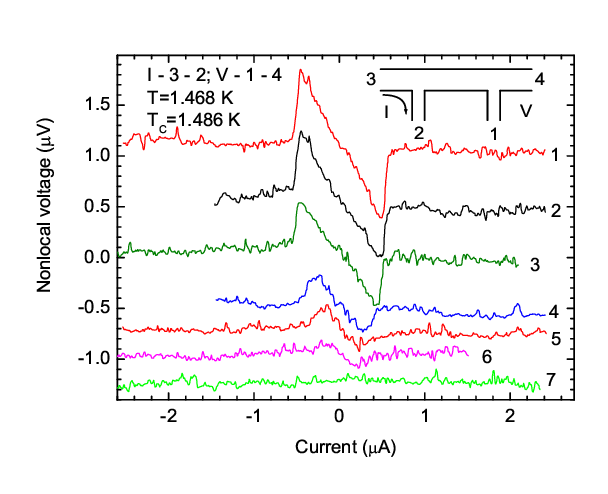}
\caption{\label{f8} (Color online) Measured nonlocal $V_{NL}(I)$
curves ($I$ - 3 - 2; $V$ - 1 - 4) at $T=1.468$\,K in different
fields $B$ for lines 1 - 7. Line 1 - $B=0$, 2 - 8.3, 3 - 12.3, 4 -
20.6, 5 - 25.7, 6 - 30.0, 7 - 48.4 G. Lines 1-2 and 4-7 are
displaced up and down vertically relative to line 3, respectively.
Inset: a sketch of the structure. The arrow indicates the
direction of the applied direct current.}
\end{center}
\end{figure}

Further, we demonstrate the nonlocal $V_{NL}(I)$ curves, recoding
in the magnetic field perpendicular to the substrate surface at
$T=1.468$\,K (Fig. \ref{f8}). Curves $V_{NL}(I)$ are no hysteresis
depending on the direction of the current sweep. The inset of Fig.
\ref{f8} shows a sketch of the structure. The voltage $V_{NL}(I)$
is measured between probes 1 and 4, located 6.69 $\mu$m from the
current circuit, when current is run through probes 3 and 2. Lines
1-2 and 4-7 are displaced up and down relative to line 3,
respectively. It is seen that the field reduces the nonlocal
voltage.

\subsection{Nonlocal critical current, negative resistance and voltage as functions of the field at $T_{1}=1.471$ and $T_{2}=1.468$\,K}

\begin{figure}
\begin{center}
\includegraphics[width=1\linewidth]{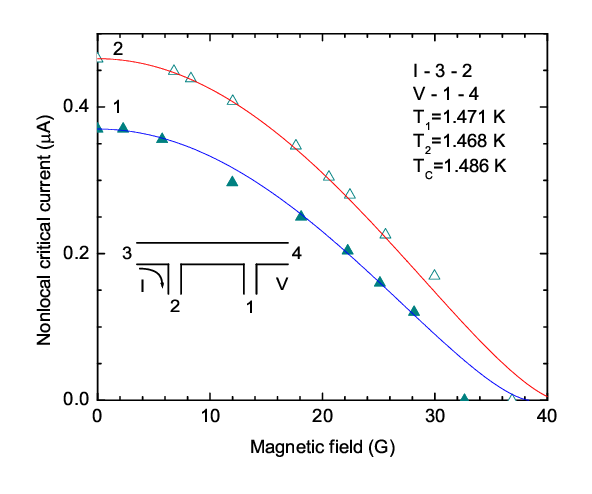}
\caption{\label{f9} (Color online) Experimental nonlocal critical
current as a function of the field $B$ at $T_{1}=1.471$\,K (closed
triangles) and $T_{2}=1.468$\,K (open triangles). Lines 1 and 2
are the fits of the nonlocal critical current (closed and open
triangles) by the functions $I_{cNL1}(T_{1},B)$ and
$I_{cNL2}(T_{2},B)$, respectively. Inset: a sketch of the
structure.}
\end{center}
\end{figure}

\begin{figure}
\begin{center}
\includegraphics[width=1\linewidth]{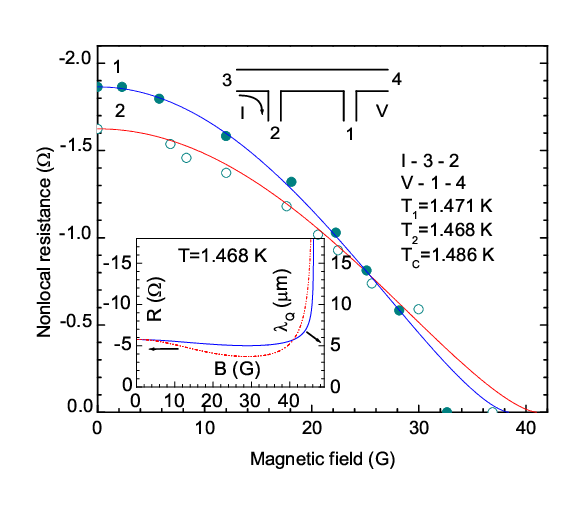}
\caption{\label{f10} (Color online) Measured negative nonlocal
resistance as a function of the field $B$ at $T_{1}=1.471 $\,K
(closed circles) and $T_{2}=1.468 $\,K (open circles). Lines 1 and
2 are the fits of negative nonlocal resistance (closed and open
circles) by the functions $R_{NL1}(T_{1},B)$ and
$R_{NL2}(T_{2},B)$, respectively. Inset upper: a sketch of the
structure. Inset below: dash-dotted and solid lines are
magnetic-field dependences of the resistance $R(x_{0},T_{2},B)$
and the quasiparticle diffusion length $\lambda_{Q}(T_{2},B)$ at
$T_{2}$, calculated within the framework of the nonequilibrium SBT
model, respectively.}
\end{center}
\end{figure}

\begin{figure}
\begin{center}
\includegraphics[width=1\linewidth]{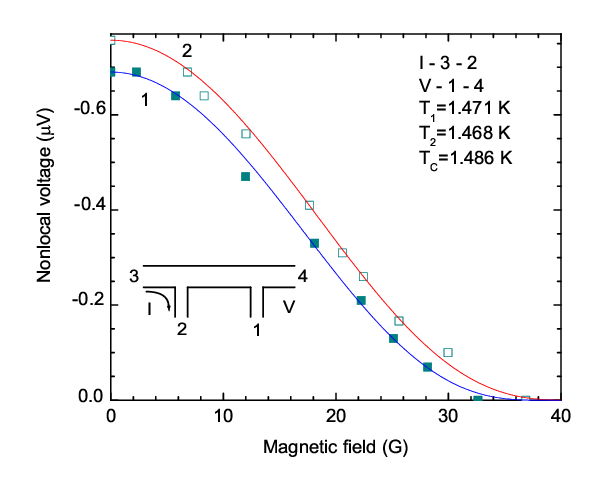}
\caption{\label{f11} (Color online) Measured maximum (peak)
negative nonlocal voltage as a function of the field $B$ at
$T_{1}=1.471$\,K (closed squares) and $T_{2}=1.468$\,K (open
squares). Lines 1 and 2 are fits of negative nonlocal voltage
(closed and open squares) with functions
$V_{NL1}(T_{1},B)=I_{cNL1}(T_{1},B)R_{NL1}(T_{1},B)$ and
$V_{NL2}(T_{2},B)=I_{cNL2}(T_{2},B)R_{NL2}(T_{2},B)$,
respectively. Insert below: a sketch of the structure.}
\end{center}
\end{figure}

Further, we present the measured and theoretical magnetic-field
dependences of nonlocal critical current, resistance and voltage
corresponding to the peak of nonlocal voltage for two values of
$T$ (Figs. \ref{f9}-\ref{f11}).

Figure \ref{f9} shows the recorded magnetic-field dependences of
the nonlocal critical current (the current coordinate of the peak
of nonlocal voltage) at $T_{1}=1.471$\,K (closed triangles) and
$T_{2}=1.468$\,K (open triangles). Lines 1 and 2 are the fits of
the nonlocal critical current for $T_{1}$ and $T_{2}$,
respectively. The sketch of the structure is shown in the inset of
Fig. \ref{f9}.

The GL critical current density $j_{GL}(T,B)$ in a
quasi-one-dimensional superconducting wire with the width $w$,
placed in a magnetic field $B$, perpendicular to the substrate
surface, at $T$ slightly below $T_{c}$, can be easily calculated
within the framework of the GL theory by minimizing the free
energy density \cite{tinkham}. Near $T_{c}$,
$j_{GL}(T,B)=j_{GL}(T,0)(1-B^{2}/B_{c}^{2}(T))^{3/2}$, where
$j_{GL}(T,0)$ is the GL critical current density at the
temperature $T$ in $B=0$ and $B_{c}(T)=\sqrt{3}\Phi_{0}/\pi
w\xi_{GL}(T)$ is the temperature-dependent maximum critical field.

Using the expression for $j_{GL}(T,B)$, we plotted fitting
theoretical magnetic-field dependences of the GL critical current
$I_{GL}(T,B)$ for two temperatures $T_{1}$ (line 1) and $T_{2}$
(line 2). Let us repeat our statement that in the temperature
range $1.455<T<1.491$ K, for any measurement circuits, the
critical current of our N-S structure is nonlocal and depends on
the region located on both sides of the place where the narrow
normal wire  connects to the wide superconducting wire. However,
in the case of Fig. \ref{f9}, the magnetic-field dependent
critical current of the structure is determined mainly by a
segment of a narrow wire with the width $w_{n}$, in which
superconductivity was induced. This is due to the fact that the
maximum critical field is inversely proportional to the wire width
($B_{c}(T) \propto 1/w$)  and hence, the superconducting order
parameter is suppressed by the magnetic field much stronger in a
wide wire than the induced superconducting order parameter in a
narrow wire.

Line 1 represents the function
$I_{cNL1}(T_{1},B)=I_{cf1}(0)(1-(B\xi_{f1}w_{n}\pi/\sqrt{3}\Phi_{0})^{2})^{3/2}$
(where $I_{cf1}(0)=0.37$ $\mu$A is the fitting critical current at
$B=0$, $\xi_{f1}=1.1$ $\mu$m is the fitting coherence length,
$w_{n}=0.27$ $\mu$m is the width of a narrow wire). Line 2 is
given by the function
$I_{cNL2}(T_{2},B)=I_{cf2}(0)(1-(B\xi_{f2}w_{n}\pi/\sqrt{3}\Phi_{0})^{2})^{3/2}$,
where $I_{cf2}(0)=0.47$ $\mu$A and $\xi_{f2}=1.03$ $\mu$m.

We found that fitting $I_{cf1}(0)$ and $I_{cf2}(0)$ almost
coincide with the measured values. However, the fitting
$I_{cf1}(0)$ and $I_{cf2}(0)$ are much lower than the
corresponding values of the GL depairing critical current in the
zero field $I_{GL}(T_{1},0)$ and $I_{GL}(T_{2},0)$, respectively.
The fitting coherence lengths $\xi_{f1}$ and $\xi_{f2}$ are close
to the corresponding calculated values $\xi_{th1}(T_{1})=1.05$ and
$\xi_{th2}(T_{2})=0.95$ $\mu$m.

Thus, the nonlocal critical current as a function of the field
(Fig. \ref{f9}) behaves like the GL depairing critical current as
a function of the field $I_{GL}(T,B)$. At the same time, the value
of the nonlocal critical current in the zero field is lower than
$I_{GL}(T,0)$.

Figure \ref{f10}, which continues the series of Figs.
\ref{f9}-\ref{f11}, shows the experimental magnetic-field
dependences of the negative nonlocal resistance corresponding to
the peak of the nonlocal voltage, at $T_{1}=1.471$\,K (closed
circles) and $T_{2}=1.468$\,K (open circles). Lines 1 and 2 are
the fits of the nonlocal resistance for $T_{1}$ and $T_{2}$,
respectively. The upper inset of Fig. \ref{f10} demonstrates a
sketch of the structure.

Line 1 in Fig. \ref{f10} is determined by the expression
$R_{NL1}(T_{1},B)=R_{f1}(0)(1-(B\xi_{f1}w_{n}\pi/\sqrt{3}\Phi_{0})^{2})^{3/2}$,
where the fitting zero-field resistance in the is
$R_{f1}(0)=-1.87$ $\Omega$ and the coherence length $\xi_{f1}=1.1$
$\mu$m. Line 2 in Fig. \ref{f10} represents a similar expression
$R_{NL2}(T_{2},B)=R_{f2}(0)(1-(B\xi_{f2}w_{n}\pi/\sqrt{3}\Phi_{0})^{2})^{3/2}$,
where $R_{f2}(0) =-1.62$ $\Omega$ and $\xi_{f2}=1.03$ $\mu$m.
Fitting $R_{f1}(0)$ and $R_{f2}(0)$ coincide with the experimental
values. Fitting coherence lengths $\xi_{f1}$ and $\xi_{f2}$ are
equal to the corresponding fitting values for the functions
$I_{cNL1}(T_{1},B)$ and $I_{cNL2}(T_{2},B)$ (Fig. \ref{f9}).

To our surprise, it turned out that the fitting functions
$R_{NL1}(T_{1},B)$ and $R_{NL2}(T_{2},B)$ are directly
proportional to the corresponding fitting functions for nonlocal
critical currents $I_{cNL1}(T_{1},B)$ and $I_{cNL2}(T_{2},B)$.
Thus, $R_{NL1}(T_{1},B)=k_{1}I_{cNL1}(T_{1},B)$, where
$k_{1}=-5.04$ $\Omega$/$\mu$A and
$R_{NL2}(T_{2},B)=k_{2}I_{cNL2}(T_{2},B)$, where $k_{2}=-3.49$
$\Omega$/$\mu$A.

Probably, in a magnetic field, the negative resistance is directly
proportional to the maximum quasiparticle current, which is equal
to the nonlocal critical current. In this case, the nonlocal
critical current is directly proportional to the GL depairing
critical current $I_{GL}(T,B)$ (Fig. \ref{f9}).

According to the nonequilibrium SBT model \cite{sbt}, the negative
nonlocal and local resistances in our structure as functions of
$B$ at $T_{2}=1.468$\,K are given by the expression
$R(x,T_{2},B)=-\lambda_{Q}(T_{2},B)\rho_{n}A^{-1}(1-tanh(x/\lambda_{Q}(T_{2},B)))$,
where $x=x_{0}=6.69$ $\mu$m is the distance between probes 1 and 2
(see subsection \ref{subsection34}). The lower inset of Fig.
\ref{f10} shows the theoretical magnetic-field dependences of
negative resistance $R(x_{0},T_{2},B)$ and the quasiparticle
diffusion length $\lambda_{Q}(T_{2},B)$ at $T_{2}=1.468$\,K.

It is seen that the theoretical and experimental resistances are
close in order of a magnitude in fields of $0-40$\,G. In
particular, in the zero field, the expected resistance is
$R(x_{0},T_{2},0)=-5.78$ $\Omega$. However, in fields close to the
critical field $B_{c}(T_{2})$, the calculated resistance
$R(x_{0},T_{2},B)$ is radically different from the measured
resistance.

Figure \ref{f11}, completing the series of Figs.
\ref{f9}-\ref{f11}, demonstrates the recorded magnetic-field
dependences of the negative nonlocal voltage corresponding to the
peak of the nonlocal voltage, at $T_{1}=1.471$\,K (closed squares)
and $T_{2}=1.468$\,K (open squares). Lines 1 and 2 are the fits of
the nonlocal voltages for $T_{1}$ and $T_{2}$, respectively. The
inset of Fig. \ref{f11} shows a sketch of the structure.

Line 1 represents the expression $V_{NL1}(T_{1},B)$, which is the
multiplication of the corresponding nonlocal critical current
$I_{cNL1}(T_{1},B)$ (Fig. \ref{f9}) and the negative resistance
$R_{NL1}(T_{1},B)$ (Fig. \ref{f10}). Then $V_{NL1}(T_{1},B)
=V_{f1}(0)(1-(B\xi_{f1}w_{n}\pi/\sqrt{3}\Phi_{0})^{2})^{3}$, where
$V_{f1}(0)=I_{cf1}(0)R_{f1}(0)=-0.69$ $\mu$V.

Similarly, line 2 is given by the function
$V_{NL2}(T_{2},B)=V_{f2}(0)(1-(B\xi_{f2}w_{n}\pi/\sqrt{3}\Phi_{0})^{2})^{3}$,
where $V_{f2}(0)=I_{cf2}(0)R_{f2}(0)=-0.76$ $\mu$V. The fitting
coherence lengths $\xi_{f1}$ and $\xi_{f2}$ are equal to the
corresponding fitting values for the functions $I_{cNL1}(T_{1},B)$
and $I_{cNL2}(T_{2},B)$ (Fig. \ref{f9}).

\section{CONCLUSION}

Earlier, in \cite{tid}, a negative nonlocal voltage is recorded by
normal and superconducting whiskers in the part of the
superconducting whisker through which no current flows, in a
temperature range of 6 mK length. In \cite{tid}, the quasiparticle
charge imbalance occurs in the phase-slip center, while in our
structure, the quasiparticle charge imbalance appears due to the
N-S interface. In \cite{tid}, to explain the negative voltage, an
unusual spatial dependence of the quasiparticle potential
$\bar{\mu}_{q}$ was proposed, which exceeds the potential of the
Cooper pairs $\bar{\mu}_{p}$ at the large distance from the core
of the phase-slip center. Whereas in the generally accepted model
\cite{sbt} $\bar{\mu}_{q}$ and $\bar{\mu}_{p}$, which differ in
the nonequilibrium region, coincide far beyond this region.

In \cite{yu}, the negative differential resistance $R=dV/dI$ is
measured on a structure consisting of superconductors S and W with
different critical currents $I_{cS}$ and $I_{cW}$. The voltage
\cite{yu} was recorded between the normal and superconducting
probes connected to the superconductor S, when the current $I$,
satisfying the requirement $I_{cW}\leqslant I\leqslant I_{cS}$,
flowed through the S-W interface. In \cite{yu}, negative
resistance appeared in a nonequilibrium case at $I\geqslant
I_{cW}$ due to the formation of the phase-slip center. Whereas in
our structure, a negative voltage (resistance) occurs in an almost
equilibrium case with a current varying from zero to $I_{c}$.

One of the important results of our work is the finding of a
decrease in the critical temperature
$dT_{c}=T_{cw}-T_{cn}=-40$\,mK of a quasi-one-dimensional
superconducting aluminum wire with a thickness of $d=19$ nm with a
decrease in the wire width from $w_{w}=0.5$ to $w_{n}=0.27$
$\mu$m.

It is well known \cite{chubov, borisenko} that $T_{c}$ of aluminum
films with a thickness $d=2\times 10^{4}-1.6$\,nm exceeds the
critical temperature of the bulk superconductor $T_{cb}=1.194$\,K.
In \cite{chubov}, the relative increase in the critical
temperature of a thin aluminum film $dT_{c}/T_{cb} \propto
S_{a}/V_{s} \propto 1/d$, reaching a value equal to one, is
measured (where $S_{a}$ and $V_{s}$ are the surface area and the
volume of the superconductor, respectively). In the work
\cite{chubov}, the films under study were about 11 mm long and
0.5-1.5 mm wide.

In \cite{borisenko}, an electron transport in a thin-film aluminum
superconducting chain consisting of rhombuses with a side of 1
$\mu$m, connected by narrow short isthmuses with a width and
length of shorter than 0.1 $\mu$m, was investigated. It was found
that the critical temperatures of a rhombus and a isthmus are
$T_{cr}=1.3$\,K and $T_{ci}=5.8$\,K, respectively.

Based on the above, we assumed that the ratio $dT_{c}/T_{cb}
\propto 1/wd$ is valid for our quasi-one-dimensional
superconducting wires. Since the ratio $w/d>10$, it should expect
a slight increase in the critical temperature with decreasing wire
width. Our additional studies have shown that the expression
$dT_{c}/T_{cb} \propto 1/d$ is valid for quasi-one-dimensional
aluminum wires with the same width $w<2\xi(T)$. However, the
expression $ dT_{c}/T_{cb} \propto 1/w$ does not hold for our
quasi-one-dimensional wires of the same width $d$. This work
indicates an inverse relationship - a decrease in $T_{c}$ with a
decrease in $w$ of the wire.

Let us speculate why the critical temperature of the narrow wire
$T_{cn}$ is lower than the critical temperature of the wider wire
$T_{cw}$. In our case, aluminum wires made by thermal deposition
on silicon substrates using the lift-off process of electron beam
lithography are dirty quasi-one-dimensional superconductors. In
such event, it should be expected that the density of impurities
and defects will be much higher at the lateral boundaries of the
wire than at the inner part of the wire. Centers are located at
the lateral boundaries of the wire, which destroy superconducting
pairs. These depairing centers can be magnetic atoms or vacancies
on which electrons land, leading to uncompensated electron spin.
Since the width of the wires is $w<2\xi(T)$, dirty lateral
boundaries of the wire cause a slight decrease in the effective
critical temperature. Most likely, the influence of the boundaries
leads to a slightly larger decrease in the effective critical
temperature in a narrower wire than this decrease in the critical
temperature in a wide wire. Thus, the condition $T_{cn}<T_{cw}$ is
satisfied. We believe that the expression $dT_{c}/T_{cb} \propto
1/w$ will be valid for superconducting aluminum films with the
width of $w>20\xi(T)$.

Thus, we have measured the nonlocal $V_{NL}(I)$ and local
$V_{L}(I)$ voltages in an aluminum superconducting
quasi-one-dimensional structure, biased with a direct current $I$,
and placed in a field $B$, perpendicular to the substrate surface,
at $T$ close to $T_{c}$. Both nonlocal $V_{NL}(I)$ and local
$V_{L}(I)$ curves showed negative voltages (resistances) when $I$
changed from zero to a critical current $I_{c}(T,B)$. With a
further increase in current, both voltages at first tend to zero,
then the local voltage changes polarity and increases linearly
with the current, while the nonlocal voltage reaches a constant
close to zero. Negative voltages appear in the range
$T_{cn}<T<T_{cw}$, where $T_{cn}$ and $T_{cw}$ are the critical
temperatures of narrow and wide wires that compose a heterogeneous
N-S structure.

Negative voltages appear when the current-carrying part of the
structure includes superconducting and normal wires, and the
voltage is taken by normal and superconducting wires. This effect
is due to the nonlocal and local quasiparticle charge imbalance
appearing due to the injection of quasiparticles from a normal
wire into a superconducting wire. Negative voltages correspond to
the negative difference between the electrochemical potentials of
quasiparticles and pairs.

For a detailed analysis, we have plotted experimental and fitting
temperature and magnetic-field dependences of the current and
potential coordinates of the peaks of negative nonlocal
$V_{NL}(I)$ and local $V_{L}(I)$ voltages. It is found that the
current coordinate of the voltage peak coincides with the critical
current for both nonlocal and local cases. In addition, we have
presented the measured and fitting temperature and magnetic-field
dependences of negative nonlocal and local resistances
corresponding to the voltage peaks. We have fitted the temperature
dependences of negative resistances using the theory of
superconducting fluctuations at $T>T_{cn}$ \cite{larkin}. We have
also calculated the temperature and magnetic-field dependences of
the resistances using the nonequilibrium SBT model at $T<T_{cw}$
\cite{sbt}.

The equilibrium and nonequilibrium models qualitatively describe
the temperature dependences of the resistances at $T$ not very
close to $T_{cw}$. To our surprise, near $T_{cw}$, the negative
nonlocal resistance is directly proportional to the critical
current and becomes zero at $T=T_{cw}$. The nonequilibrium model
qualitatively describes the magnetic-field dependence of negative
resistance at a given $T$ in fields many weaker than the
temperature-dependent maximum critical field $B_{c}(T)$. It is
unexpectedly found that the negative nonlocal resistance as a
function of the field is directly proportional to the GL depairing
critical current $I_{GL}(T,B)$.

Thus, for a more adequate theoretical understanding of the
experimental results, a model is needed that takes into account
the simultaneous presence of equilibrium and nonequilibrium
superconducting fluctuations in the N-S structure.

\section{ACKNOWLEDGMENTS}

The authors are grateful to A. Firsov for preparing the structures
and M. Skvortsov for useful discussions. The work was done within
the framework of the State Task No. 075-00920-20-00.





\end{document}